\documentclass{article}

\usepackage{arxiv}

\usepackage[utf8]{inputenc} 
\usepackage[T1]{fontenc}    
\usepackage{hyperref}       
\usepackage{url}            
\usepackage{booktabs}       
\usepackage{amsfonts}       
\usepackage{nicefrac}       
\usepackage{microtype}      
\usepackage{lipsum}		
\usepackage{physics}
\usepackage{graphicx}
\usepackage{natbib}
\usepackage{doi}

\title{Musical Molecules: Sonifying the IR Spectra and Modeling Intramolecular Vibrational Energy Redistribution of Small Molecules}

\author{ \href{https://orcid.org/0009-0003-8247-3927}{\includegraphics[scale=0.06]{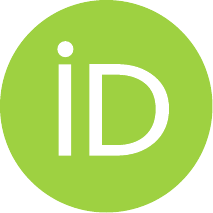}\hspace{1mm}Sophia H.~Kim} \\
	Department of Physics\\
	Harvard University\\
	Cambridge, MA 02138 \\
	\texttt{sophiakim@college.harvard.edu} \\
	\And \href{https://orcid.org/0000-0002-9907-9497}
	{\includegraphics[scale=0.06]{orcid.pdf}\hspace{1mm}Eric J.~Heller} \\
	Department of Physics\\
	Harvard University\\
	Cambridge, MA 02138 \\
	\texttt{eheller@fas.harvard.edu} \\
}

\renewcommand{\shorttitle}{Musical Molecules}

\hypersetup{
pdftitle={A template for the arxiv style},
pdfsubject={physics.chem-ph},
pdfauthor={Sophia H.~Kim, Helena H.~Kim},
pdfkeywords={IR spectra, intramolecular vibrational energy redistribution, sonification},
}

\begin{document}
\maketitle

\begin{abstract}
	\ This work explores how small molecules sound. Infrared (IR) spectra of HCl, H\textsubscript{2}O, NH\textsubscript{3}, and acetone are mapped into the audible range using a simple anharmonic oscillator model and NIST vibrational data. Comparing harmonic and anharmonic sonifications reveals systematic pitch flattening, beating, and the emergence of combination bands, which are analyzed with spectrograms and autocorrelation functions. A time-dependent model of intramolecular vibrational energy redistribution (IVR) in acetone, implemented by “plucking’’ a single mode, produces evolving sound textures that mirror energy flow through the molecule. These results suggest that sonified IR spectra can provide an intuitive, pedagogical window into anharmonicity, mode coupling, and IVR.

\end{abstract}

\keywords{IR spectra \and intramolecular vibrational energy redistribution \and sonification}

\section{Introduction}

Sonification is a well-established approach in the scientific literature for providing intuitive understanding of abstract concepts and facilitating discovery. In chemistry, Arasaki and Takasuka extracted characteristic energetic structures from complex electronic wavefunctions to sonify molecular electronic energy density \citep{Arasaki}, while others have explored translating organic molecular structures into musical compositions \citep{mahjour_molecular_2023}. Sonification also offers significant pedagogical potential. In 2013, a study sonified infrared spectra of molecules and investigated their interpretation by blind and visually impaired students \citep{Pereira}. A chemical education study mapped infrared wavenumbers onto a scale of musical notes \citep{Garrido}. 

While previous studies of IR spectral sonification have focused on creating "musical scales," there has been less emphasis on the acoustical quality and characteristics of the resulting tones. This paper builds on these earlier efforts by sonifying IR spectra and subsequently analyzing them using psychoacoustical techniques. 

This work has three primary aims: (1) create an auditory comparison between harmonic and anharmonic molecular spectra, (2) progressively increase complexity from diatomic molecules to water to polyatomic molecules, and (3) sonify intramolecular vibrational energy redistribution (IVR) in acetone. While the project explored ammonia, methanol, and acetone, this paper focuses on ammonia for the static spectral comparison and acetone for the dynamic IVR simulation. The third aim is of particular interest, as it enables one to hear the acoustic signature of "plucking" or selectively exciting a single vibrational mode and allowing the energy to propagate throughout the molecule.

\section{Methods}

\subsection{IR Spectra Data Sources}
Experimental IR spectroscopic data were obtained from the NIST Chemistry WebBook \citep{nist_webbook}. Unless otherwise noted, gas-phase spectra (typically the first table listed) were used as reference values. Spectral features can vary with molecular phase, temperature, and experimental methodology; such variations were not systematically explored in this work.

\subsection{Deriving the Anharmonic Spectrum}

We model molecular vibrations as perturbed harmonic oscillators:
\begin{equation}
    H = H_0 + V
\end{equation}
where $H_0 = \hbar\omega(a^\dagger a + \frac{1}{2})$ is the unperturbed harmonic oscillator Hamiltonian and $V = -\alpha x^3 - \beta x^4$ represents anharmonic corrections.

First-order perturbation theory gives the energy shift for state $\ket{n}$:
\begin{equation}
    \Delta E^{(1)}_n = \bra{n}V\ket{n}
\end{equation}
This follows from the standard perturbative expansion $E_n = E_n^{(0)} + \lambda\bra{n^{(0)}}V\ket{n^{(0)}} + \mathcal{O}(\lambda^2)$ \citep{anharmonic_oscillator_notes, heller_semiclassical}. Physically, the first-order energy correction equals the expectation value of the perturbation in the unperturbed eigenstate.

Expressing position in ladder operators, $x = \sqrt{\frac{\hbar}{2m\omega}}(a + a^\dagger)$, we expand the potential terms. The cubic term:
\begin{equation}
    x^3 = \left(\frac{\hbar}{2m\omega}\right)^{3/2} (a + a^\dagger)^3
\end{equation}
yields $\bra{n}x^3\ket{n} = 0$ because $(a + a^\dagger)^3$ contains only terms connecting states with $\Delta n = \pm 1, \pm 3$ (odd parity), so all diagonal matrix elements vanish.

The quartic term dominates:
\begin{equation}
    \bra{n}x^4\ket{n} = \left(\frac{\hbar}{2m\omega}\right)^2 \bra{n}(a + a^\dagger)^4\ket{n} = \left(\frac{\hbar}{2m\omega}\right)^2 (6n^2 + 6n + 3)
\end{equation}
This result uses the commutation relation $[a, a^\dagger] = 1$ and the diagonal selection rule: only terms with equal numbers of creation and annihilation operators contribute. Explicitly, the surviving terms from $(a + a^\dagger)^4$ are $a a^\dagger a a^\dagger$, $a^\dagger a a^\dagger a$, $a^\dagger a^\dagger a a$, etc., which sum to $6n^2 + 6n + 3$ when evaluated in state $\ket{n}$.

Then, we can rewrite $6n^2 + 6n + 3 = 6(n + \frac{1}{2})^2 + \tfrac{3}{2}$, and then by dropping the constant shift (which cancels in transition energies):
\begin{equation}
    E_n \approx \hbar\omega\left(n + \frac{1}{2}\right) - \beta\left(\frac{\hbar}{2m\omega}\right)^2 \cdot 6\left(n + \frac{1}{2}\right)^2
\end{equation}

Defining the anharmonicity constant $\chi$:
\begin{equation}
    \chi \equiv \frac{6\beta\hbar^2}{4m^2\omega^3} = \frac{3\beta\hbar}{2m^2\omega^3}
\end{equation}
the corrected energy levels become:
\begin{equation}
    E_n = \hbar\omega\left(n + \frac{1}{2}\right)\left[1 - \chi\left(n + \frac{1}{2}\right)\right]
\end{equation}

For transitions from the ground state ($n = 0$) to level $n$:
\begin{equation}\label{eq:anharmonic}
    f_n = \frac{E_n - E_0}{h} = nf_0\left[1 - \chi(n + 1)\right]
\end{equation}
where $f_0 = \omega/(2\pi)$ is the harmonic fundamental frequency. This predicts systematic redshifting (flattening) of overtones, with the shift proportional to $\chi$ and quadratic in $n$.
\subsection{Polyatomic Spectra}

A computational challenge arises as the number of vibrational modes increases: the total number of possible overtone and combination bands grows combinatorially. It becomes inefficient to model all possible combinations, particularly since weaker combinations may be inaudible in the final sound file. While for diatomic molecules we considered integer excitations ($n = 0, 1, 2, \ldots$) of a single vibrational mode, for polyatomic molecules we focus on the fundamental transition ($n = 1$) of each mode and the various binary and higher-order combinations that can arise from mode coupling.

To maintain computational efficiency, we implemented a filtering strategy based on NIST IR intensity classifications \citep{nist_webbook}. We assigned amplitude weights according to the qualitative intensity ratings: 1.0 for VS (very strong), 0.7 for S (strong), and 0.4 for M (moderate). Modes classified as W (weak) were excluded from the simulation entirely. For combination bands, only those involving strong parent modes (VS-VS, VS-S, or S-S pairings) were retained, as weaker combinations would contribute negligibly to the audible spectrum. This filtering reduced the number of spectral features while preserving the most prominent vibrational signatures of each molecule.

\subsection{Intramolecular Vibrational Energy Redistribution}

When a specific vibrational mode of a molecule is "plucked" or selectively excited, the energy does not remain localized but propagates to other vibrational modes. This process, known as intramolecular vibrational redistribution (IVR), can exhibit energy transfer to coupled modes and potentially quantum recurrence back to the initially excited mode.

As in the anharmonic oscillator treatment, we model the system Hamiltonian using perturbation theory:

\begin{equation}
    H = H_0 + V_a
\end{equation}

where $V_a$ is the anharmonic potential that couples different modes:

\begin{equation}
    V_a = \underbrace{\sum_i k_{iii} q_i^3}_{\text{self-anharmonicity}} + \underbrace{\sum_{i<j} k_{iij} q_i^2 q_j}_{\text{two-mode coupling}} + \underbrace{\sum_{i<j<k} k_{ijk} q_i q_j q_k}_{\text{three-mode coupling}} + \underbrace{\sum_{ijkl} k_{ijkl} q_i q_j q_k q_l}_{\text{quartic terms}} + \cdots \tag{11}
\end{equation}

The presence of cubic and higher-order terms is essential for IVR. In purely harmonic systems, normal modes are independent and cannot exchange energy. While quadratic (bilinear) coupling terms $\sum_{ij} k_{ij} q_i q_j$ can be eliminated by diagonalization to yield uncoupled normal modes, nonlinear terms (cubic and higher) introduce irreducible mode coupling that enables energy redistribution.

Here $q_i$ denotes normal mode coordinates, $k_{ijk}$ represents 3-mode coupling coefficients (cubic anharmonicity), and $k_{ijkl}$ represents 4-mode coupling (quartic anharmonicity). These cross-anharmonicity terms couple different modes and correspond to the coupling matrix elements in our simulation. Specifically: self-anharmonicity $k_{iii}q_i^3$ enables transitions between energy levels within a single mode; two-mode cubic coupling $k_{iij}q_i^2q_j$ enables energy transfer between mode pairs; three-mode coupling $k_{ijk}q_iq_jq_k$ introduces additional resonances; and quartic terms $k_{ijkl}$ encode higher-order couplings.

\subsubsection{Time-Dependent Amplitude Model}

Since we are modeling impulsive excitation ("plucking"), we expect an initial decay of the excited mode followed by growth and subsequent decay of receiving modes, with possible quantum recurrence. 

The initially excited mode (mode 1, the C=O stretch in acetone) evolves as:

\begin{equation}
    A_1(t) = e^{-t/\tau} \cdot \left[1 + a\cos(2\pi t/T_{\text{rec}})\right]
\end{equation}

The exponential $e^{-t/\tau}$ captures energy decay from the plucked mode, where $\tau = 0.8~\text{s}$ is the IVR timescale. The cosine term models quantum recurrence with period $T_{\text{rec}} = 4~\text{s}$, scaled by amplitude $a = 0.2$. For small molecules like acetone, the recurrence time is estimated as $T_{\text{rec}} \sim h/\Delta E$, where $\Delta E$ is the typical energy spacing between coupled modes.

Modes that receive energy transfer (modes $i > 1$) have amplitudes:

\begin{equation}
    A_i(t) = V_{1i} \cdot g(t) \cdot d(t) \cdot b(t)
\end{equation}

where $V_{1i}$ is the coupling matrix element from mode 1 to mode $i$, and the growth function $g(t)$ is:

\begin{equation}
    g(t) = \begin{cases}
    1 - e^{-(t-\Delta_i)/\tau}, & t > \Delta_i \\
    0, & t \leq \Delta_i
    \end{cases}
\end{equation}
where $\Delta_i = 0.2(i-1)$ introduces staggered activation times for different modes. We represent the decay function $b(t)$ as:

\begin{equation}
    d(t) = e^{-\alpha(t-t_p)/\tau}
\end{equation}
where $t_p$ is the peak time and $\alpha = 0.2$ controls the decay rate. We estimate $t_p = \tau(1 + V_{1i})$ based on the physical expectation that stronger coupling leads to earlier energy transfer peaks. We represent the beating pattern $b(t)$ as:

\begin{equation}
    b(t) = \alpha_b + \beta_b\cos\left(\frac{2\pi t}{T_{\text{rec}}(1 + c \cdot i)}\right)
\end{equation}
where $\alpha_b = 0.85$, $\beta_b = 0.15$, and $c = 0.3$. This term accounts for interference between modes and introduces audible beating in the files. The mode-dependent factor $(1 + c \cdot i)$ allows different recurrence periods for each mode.

\subsubsection{Coupling Matrix and Fermi's Golden Rule}

The coupling matrix elements $V_{1i}$ were estimated based on spectroscopic selection rules and typical coupling strengths from the literature \citep{nist_webbook}. Modes with similar frequencies or matching symmetry (e.g., both CH$_3$ stretches) were assigned stronger coupling ($V_{1i} \sim 0.7-0.8$), while modes with large frequency differences or incompatible symmetries received weaker coupling ($V_{1i} \sim 0.1-0.3$). Modes with very weak coupling were excluded to reduce computational cost and because their contributions would be inaudible.

The energy transfer rate from mode 1 to mode $i$ is governed by Fermi's Golden Rule (in first-order approximation):

\begin{equation}
    \Gamma_{1\rightarrow i} = \frac{2\pi}{\hbar} \left|\bra{i}V_a\ket{1}\right|^2 \rho(E)
\end{equation}

where $\rho(E)$ is the density of states at the energy of mode 1. Higher densities of states lead to faster energy transfer and more efficient IVR. The total decay rate is $\tau^{-1} = \sum_i \Gamma_{1\rightarrow i}$, representing the sum of all energy transfer pathways from the initially excited mode.

\subsection{Sound Files and Spectra Plots}
All simulations were generated in Mathematica code. For easier viewing, there is also a  \href{https://sites.google.com/college.harvard.edu/musicalmolecules/home}{website} that one can peruse through to view all the generated sound files and images. This write-up contains commentary but it is highly recommended to check that supplementary material.

All generated tones on Mathematica were monotonic sine tones. Analysis of sound files (specifically the spectrogram and autocorrelation) were performed with Audacity \citep{audacity}.

\section{Results}

While the paper provides a written summary and some discussion of the results along with the generated Mathematica plots, it is highly suggested for the viewer to peruse through the attached website, which is perhaps more accessible and easier to \href{https://sites.google.com/college.harvard.edu/musicalmolecules/home}{navigate}. Supplementary sound files may also be accessed on that website. 

\subsection{Diatomic Molecules}

\begin{figure}[!h]\centering
  \begin{minipage}[b]{0.49\textwidth}
    \centering
    \includegraphics[width=\textwidth]{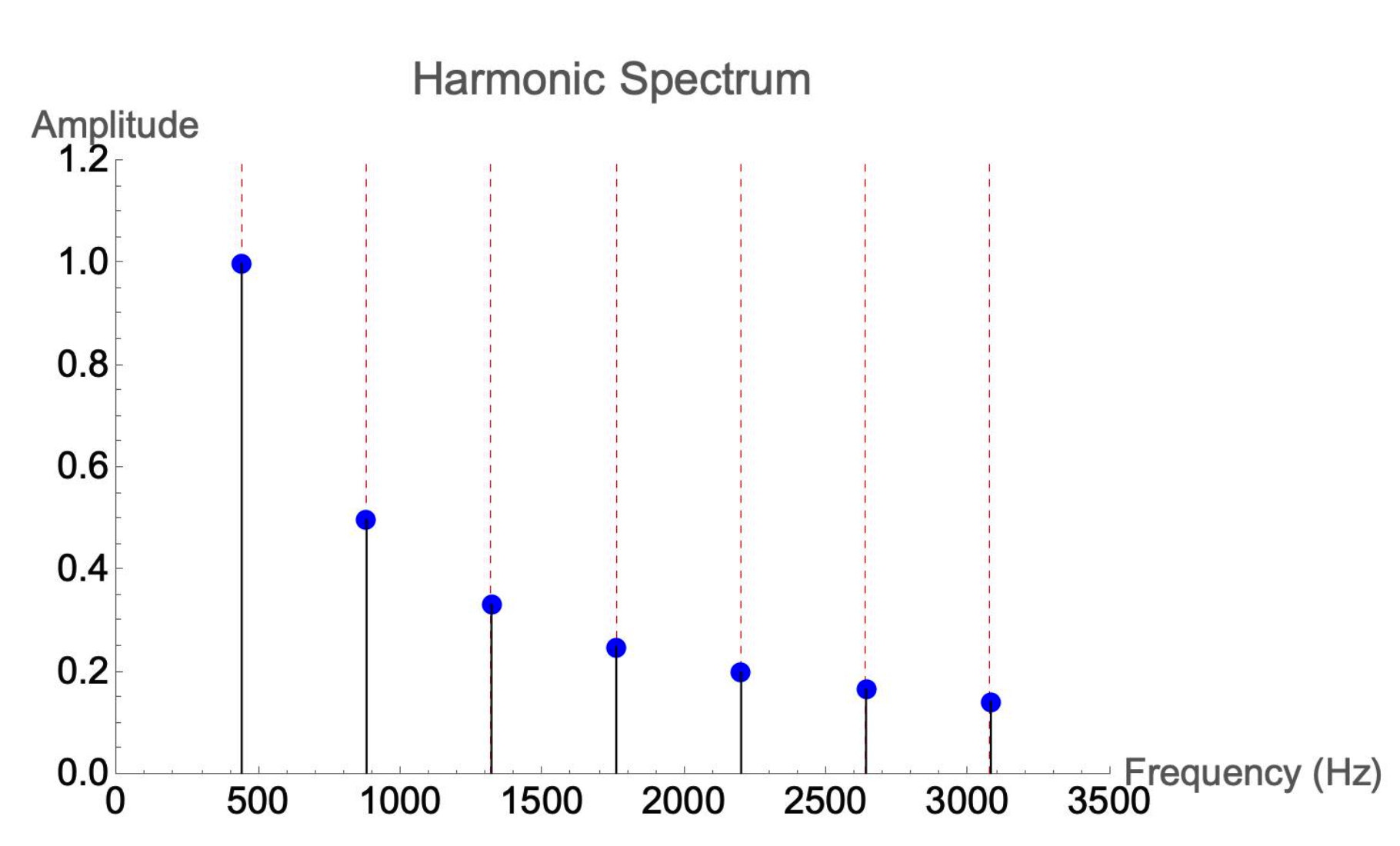}
  \end{minipage}
  \hfill
  \begin{minipage}[b]{0.49\textwidth}
    \centering
    \includegraphics[width=\textwidth]{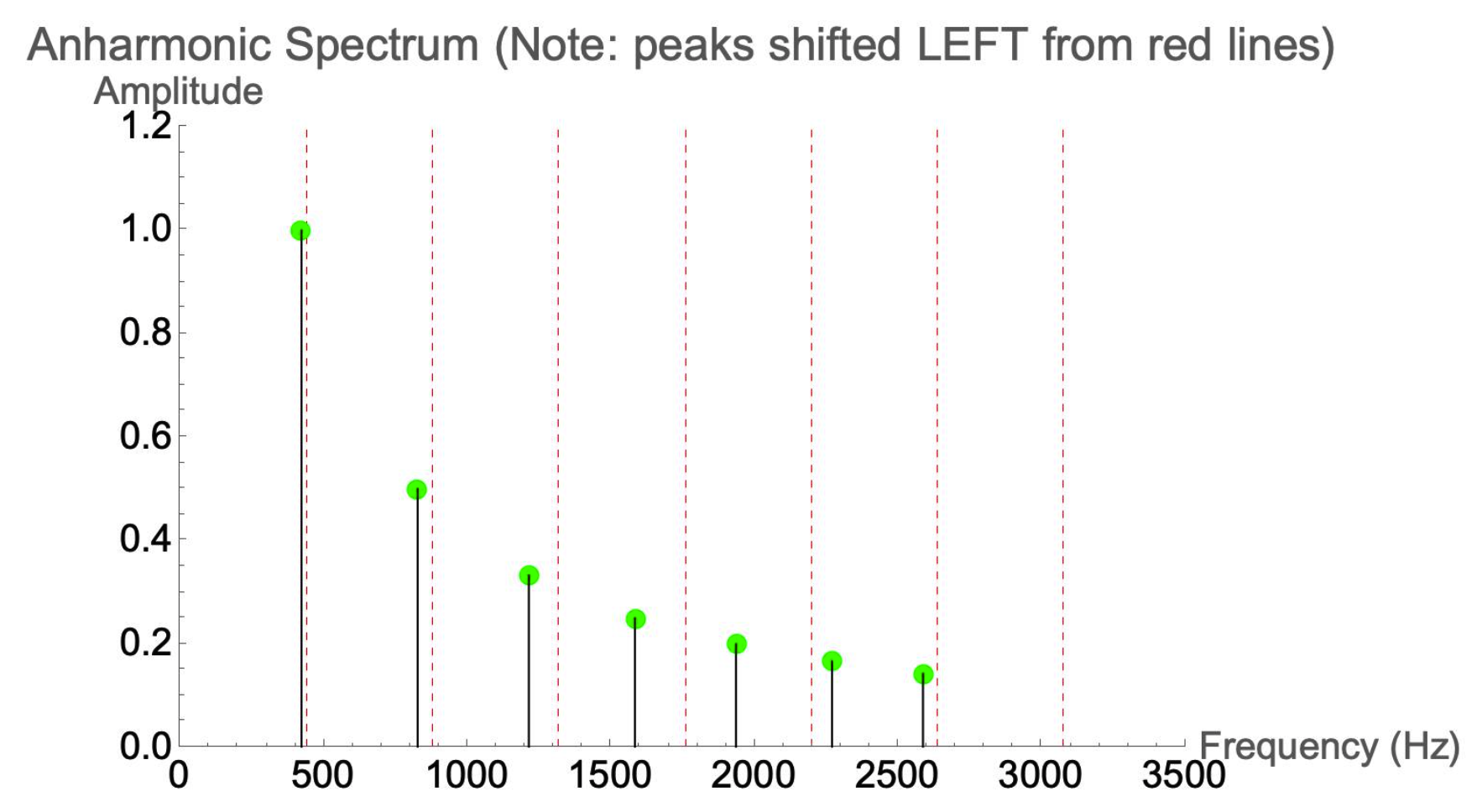}
  \end{minipage}
  \caption{Harmonic (left) and anharmonic (right) spectra of the HCl diatomic molecule, scaled from a fundamental sine frequency of 440 Hz. Note the left-shifting of peaks due to anharmonicity constant $\chi$, related to the derived equation $nf_0[1-\chi(n+1)]$.}
  \label{ch02:fig01}
\end{figure}

We base the simulation off of known wavenumber for HCl. We also use the anharmonicity constant $52.8 \text{cm}^{-1}$ \citep{nist_webbook}. Since we want to scale these wavenumbers into audible frequencies, we could choose a base fundamental frequency to choose our scaling factors. In the Mathematica notebook, I chose $f_0=440 \text{Hz}$ and generated the spectrum by multiplying the fundamental frequency by integer multiples. The reciprocal $\chi =\frac{1}{52.8}=0.02$. We calculated the anharmonic frequenices with \eqref{eq:anharmonic} and notice that the anharmonic spectrum is slightly shifted to the left, as we would expect (Fig.~\ref{ch02:fig01}).

\begin{figure}[!h]\centering
  \begin{minipage}[b]{0.8\textwidth}
    \centering
    \includegraphics[width=\textwidth]{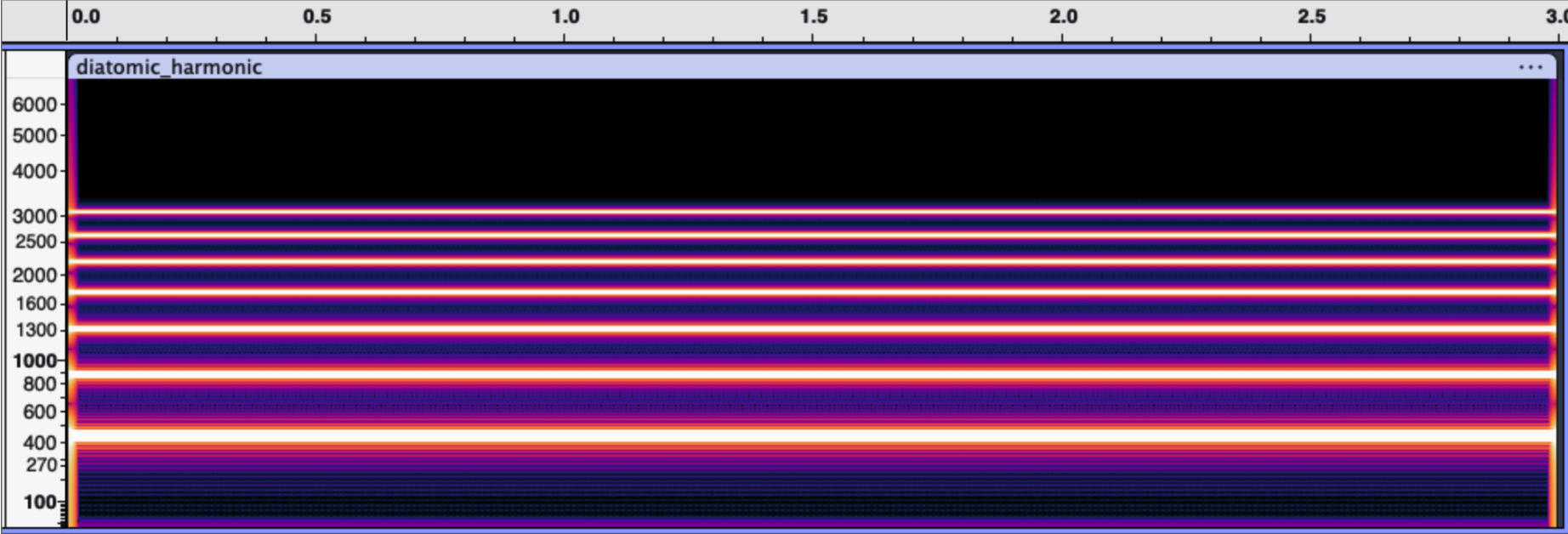}
  \end{minipage}
  \hfill
  \begin{minipage}[b]{0.8\textwidth}
    \centering
    \includegraphics[width=\textwidth]{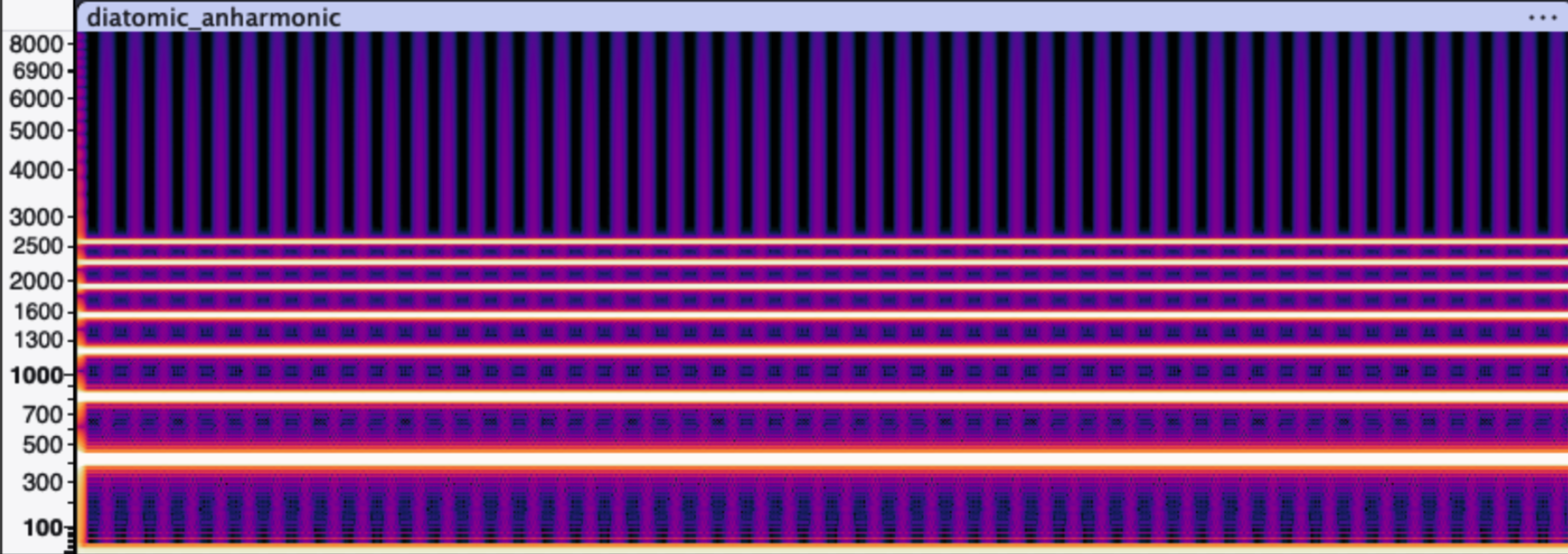}
  \end{minipage}
  \caption{Harmonic (top) and anharmonic (bottom) spectrogram of the HCl diatomic molecule, scaled from a fundamental sine frequency of 440 Hz. Note the significant beating (vertical lines) present in the anharmonic spectrogram.}
  \label{ch02:fig02}
\end{figure}

\begin{figure}[!h]\centering
  \begin{minipage}[b]{0.49\textwidth}
    \centering
    \includegraphics[width=\textwidth]{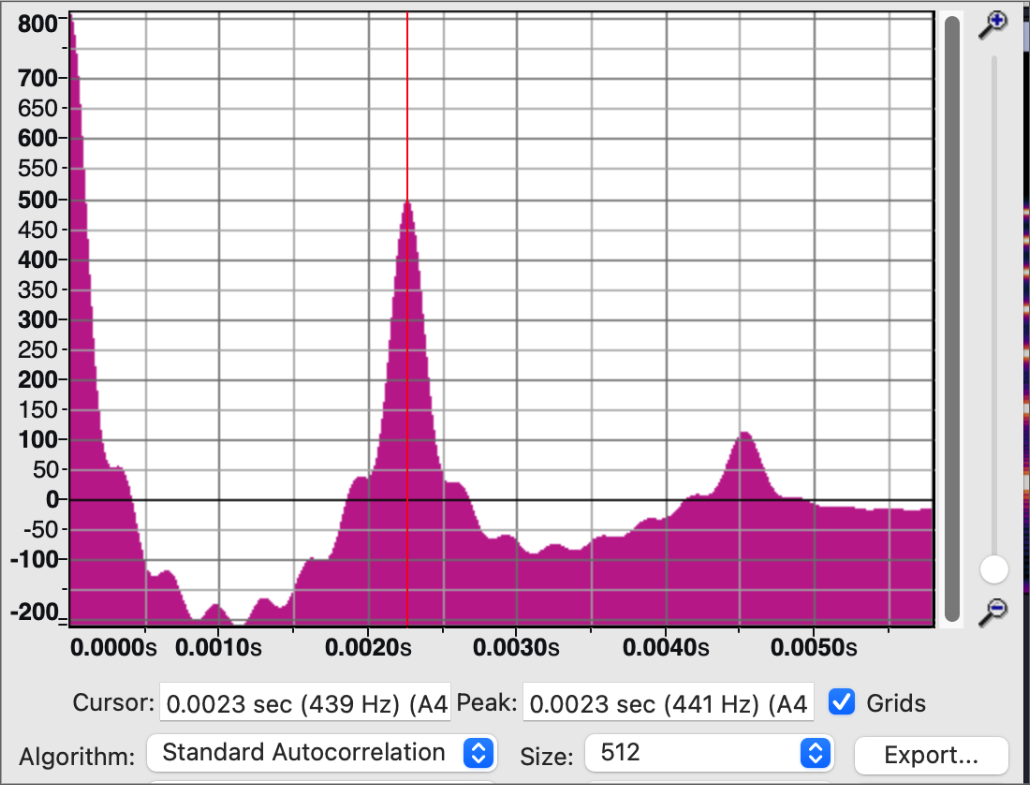}
  \end{minipage}
  \hfill
  \begin{minipage}[b]{0.47\textwidth}
    \centering
    \includegraphics[width=\textwidth]{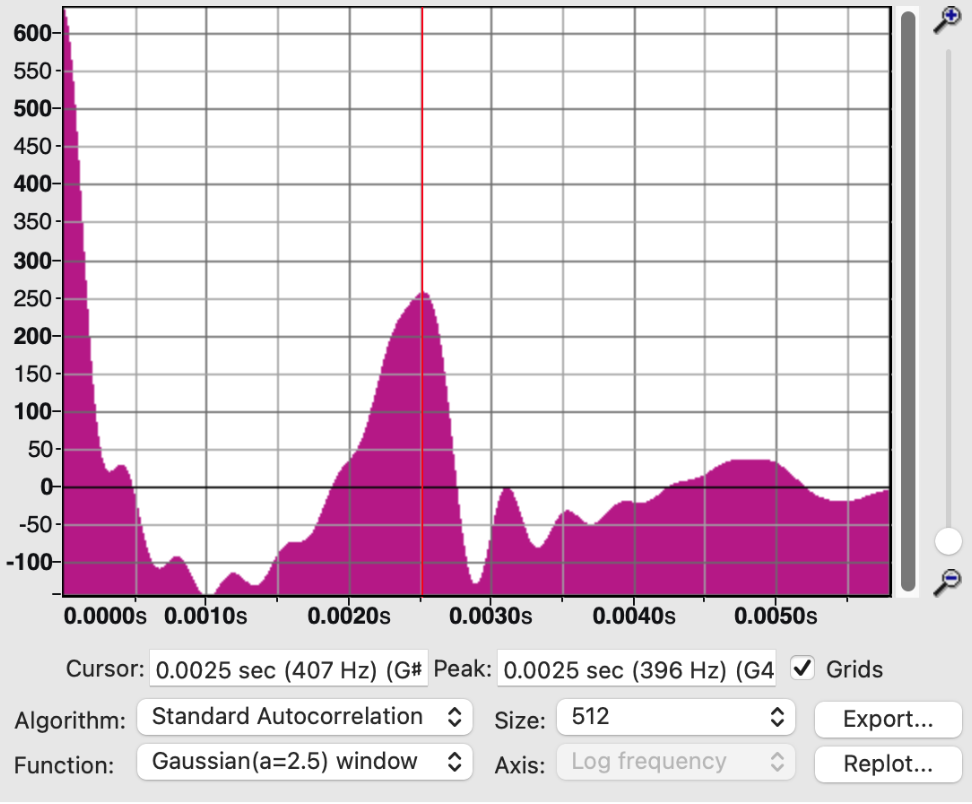}
  \end{minipage}
  \caption{Harmonic (left) and anharmonic (right) autocorrelation of the HCl diatomic molecule, scaled from a fundamental sine frequency of 440 Hz. Note the right-shifting of the first major peak that contributes to a "flatter" sound.}
  \label{ch02:fig03}
\end{figure}

Using Audacity, we also analyzed the visible spectrogram (Fig.~\ref{ch02:fig02}) and autocorrelation (Fig.~\ref{ch02:fig03}) function of the two sound files. A noticeable difference in the sonogram (Fig.~\ref{ch02:fig02}) is that the one for the anharmonic case demonstrates rapid beating with vertical lines, which could also be heard on the sound file. 

Additionally, the rightward shifting of the peak in the autocorrelation suggests the anharmonic molecule sounds "flatter" than the harmonic one (Fig.~\ref{ch02:fig03}).

\subsection{$H_2O$}

$H_2O$ has 3 vibrational modes: the symmetric, antisymmetric stretching and bending. In the literature they correspond to wavenumbers $3657 \text{cm}^{-1}$, $3756 \text{cm}^{-1}$, and $1595 \text{cm}^{-1}$, respectively \citep{nist_webbook}. The lowest wavenumber of $1595 \text{cm}^{-1}$, so we will use that as a reference point of scaling to our fundamental frequency. 

This time, we chose $220 \text{Hz}$ as our fundamental frequency. Thus, we scaled the other frequencies in Mathematica code by the ratio of the wavenumber to $1595 \text{cm}^{-1}$: 

\begin{equation}
    f_1 = f_0\cdot \frac{3657}{1595}=(220)\frac{3657}{1595} \text{Hz}
\end{equation}

and so on. We also use the anharmonicity constants $\chi_1=0.005, \chi_2=0.008, \chi_3=0.005$ where $\chi_2=0.008$ corresponds to the bending mode which is more anharmonic \citep{nist_webbook}. 

\begin{figure}[!h]\centering
  \begin{minipage}[b]{0.40\textwidth}
    \centering
    \includegraphics[width=\textwidth]{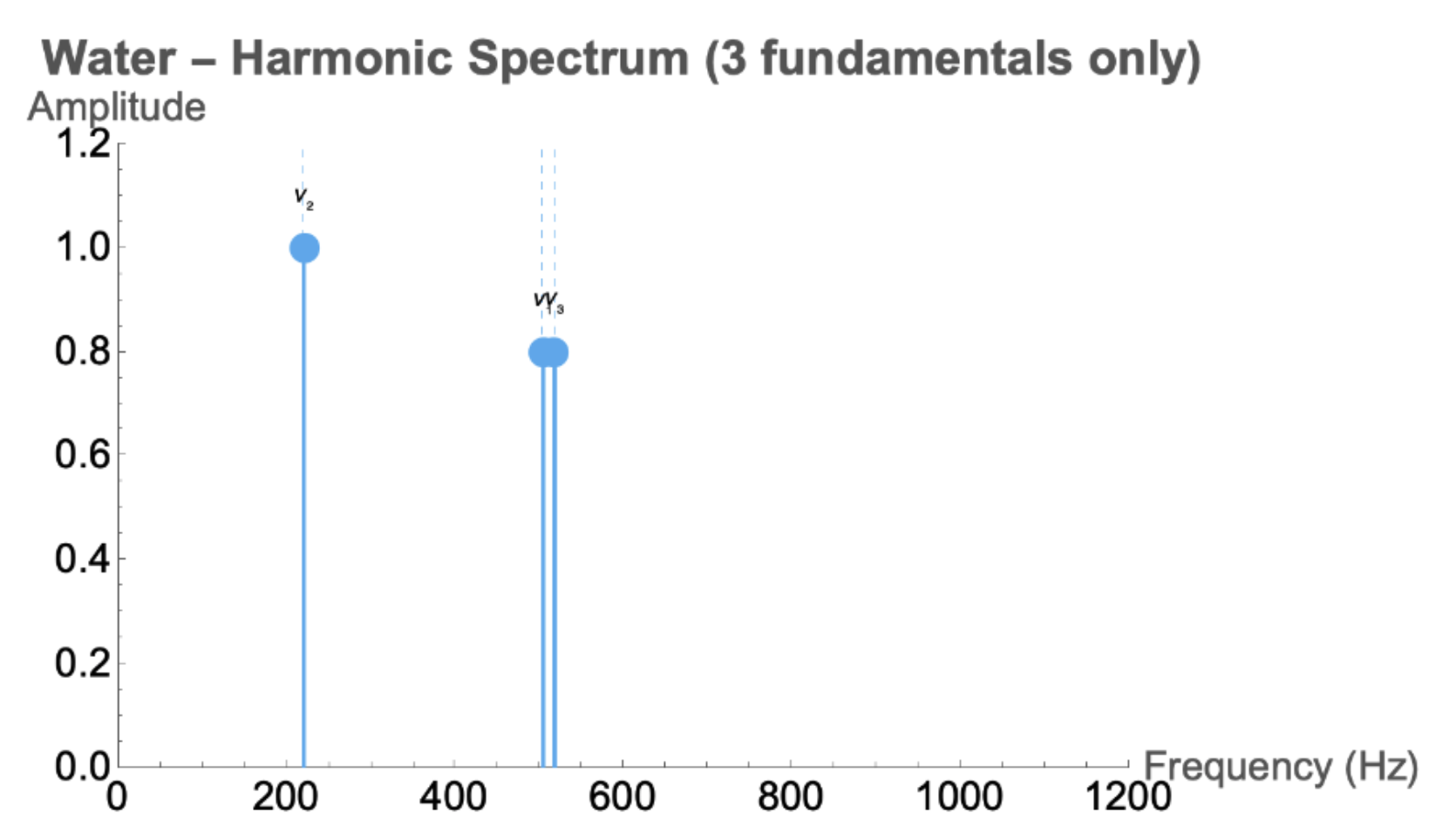}
  \end{minipage}
  \hfill
  \begin{minipage}[b]{0.60\textwidth}
    \centering
    \includegraphics[width=\textwidth]{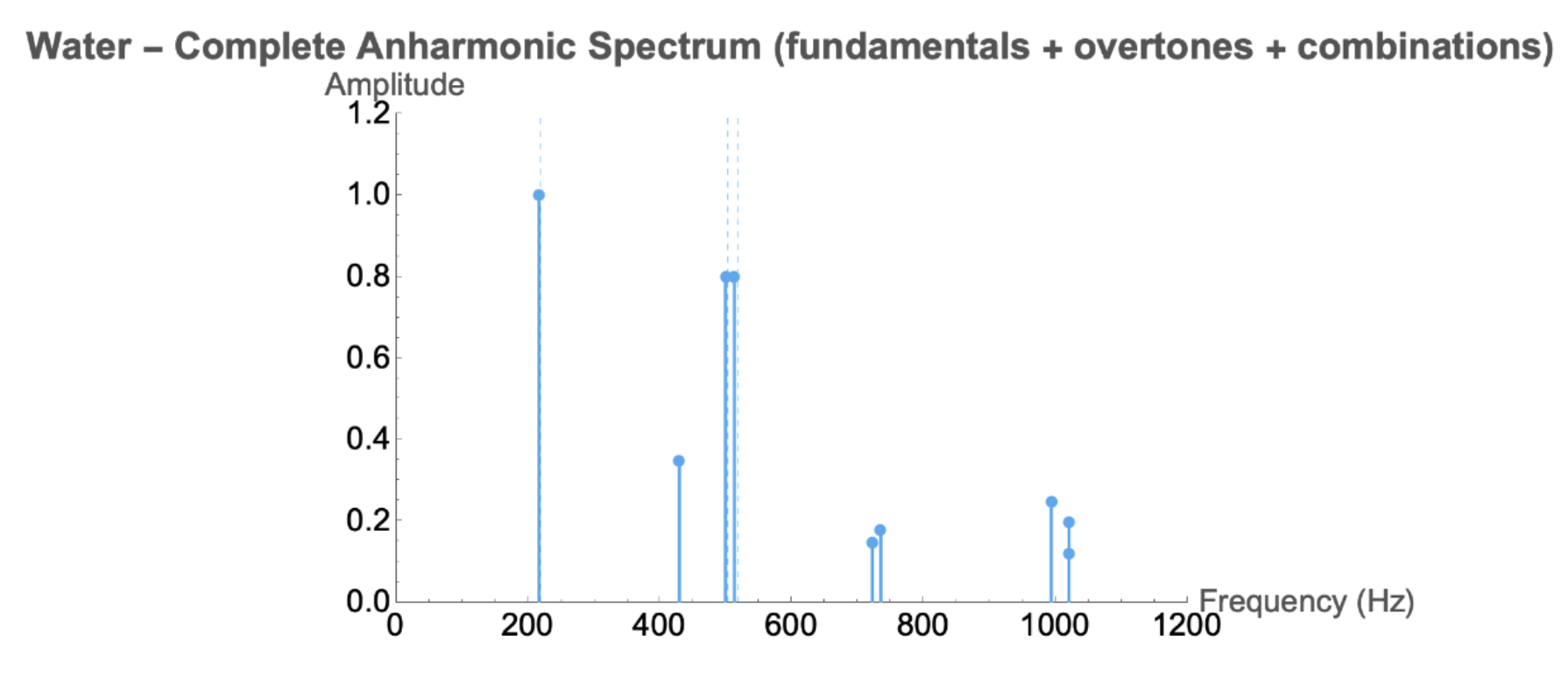}
  \end{minipage}
  \caption{Harmonic (top) and anharmonic (bottom) spectra of water, scaled from a fundamental sine frequency of 220 Hz. Note the left-shifting of peaks due to anharmonicity constant $\chi$, as well as the emergence of various combination bands of different vibrational energy modes.}
  \label{ch02:fig04}
\end{figure}

In the harmonic approximation, there are no combination bands or overtones because we have unperturbed, independent eigenstates. In contrast in the anharmonic approximation, we now have perturbed eigenstates that will interfere and we thus need to account for all the possible combinations of our frequencies. After constructing the corresponding anharmonic frequencies from the anharmonicity constants and conversions as done in \eqref{eq:anharmonic}, we use code including the complete set of combinations bands and overtones for water. We consequently developed the following spectrum displayed in Fig.~\ref{ch02:fig04}. We not only see left-shifted frequency peaks but also various new emerging variation bands from calculating the anharmonic spectrum (which is appropriately scaled to the range of human hearing).

\begin{figure}[!h]\centering
  \begin{minipage}[b]{1\textwidth}
    \centering
    \includegraphics[width=\textwidth]{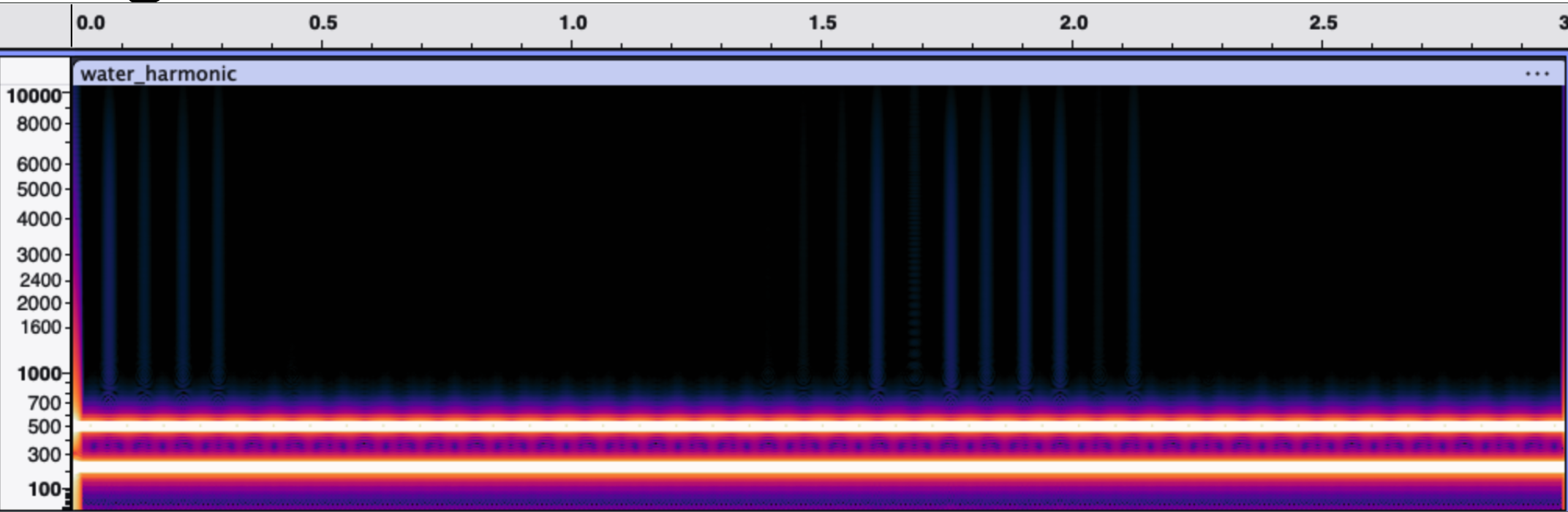}
  \end{minipage}
  \hfill
  \begin{minipage}[b]{1\textwidth}
    \centering
    \includegraphics[width=\textwidth]{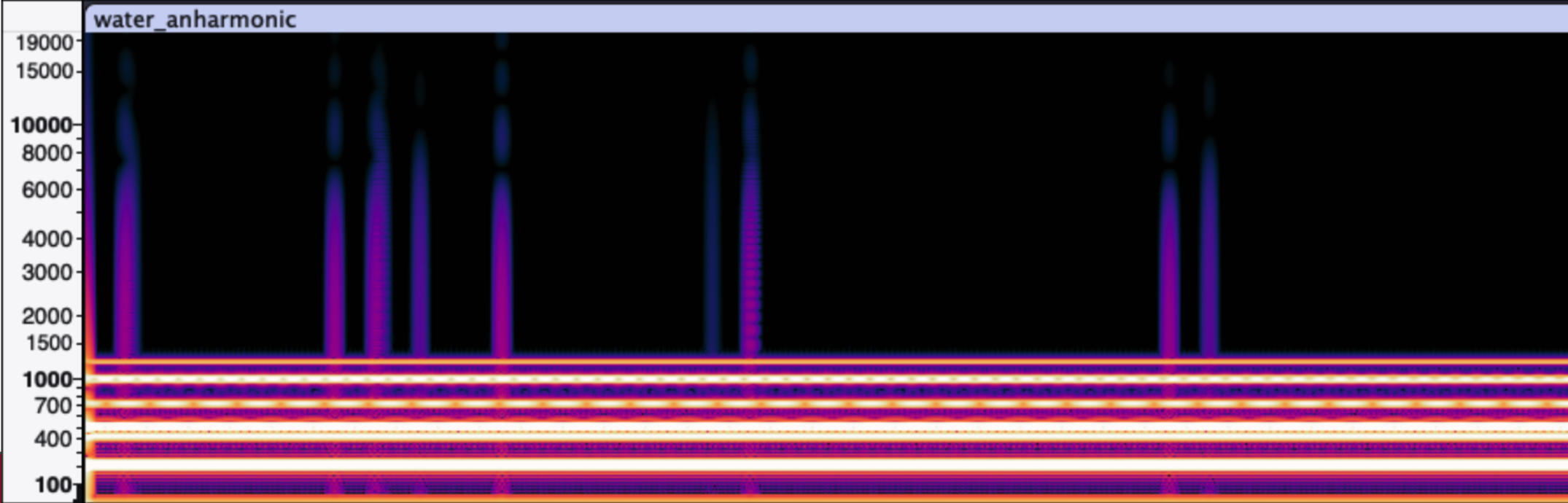}
  \end{minipage}
  \caption{Harmonic (top) and anharmonic (bottom) spectrograms of water, scaled from a fundamental sine frequency of 220 Hz. The anharmonic spectrogram shows more complexity and also brighter vertical bands than in the harmonic.}
  \label{ch02:fig05}
\end{figure}

The spectrograms (Fig.~\ref{ch02:fig05}) were quite interesting to examine, with the anharmonic spectrogram for water appearing more complex. There are also several vertical lines (though not as periodically as in HCl) that suggest occasional and more noticeable beating than in the harmonic case with fainter vertical bands. 

\begin{figure}[!h]\centering
  \begin{minipage}[b]{0.48\textwidth}
    \centering
    \includegraphics[width=\textwidth]{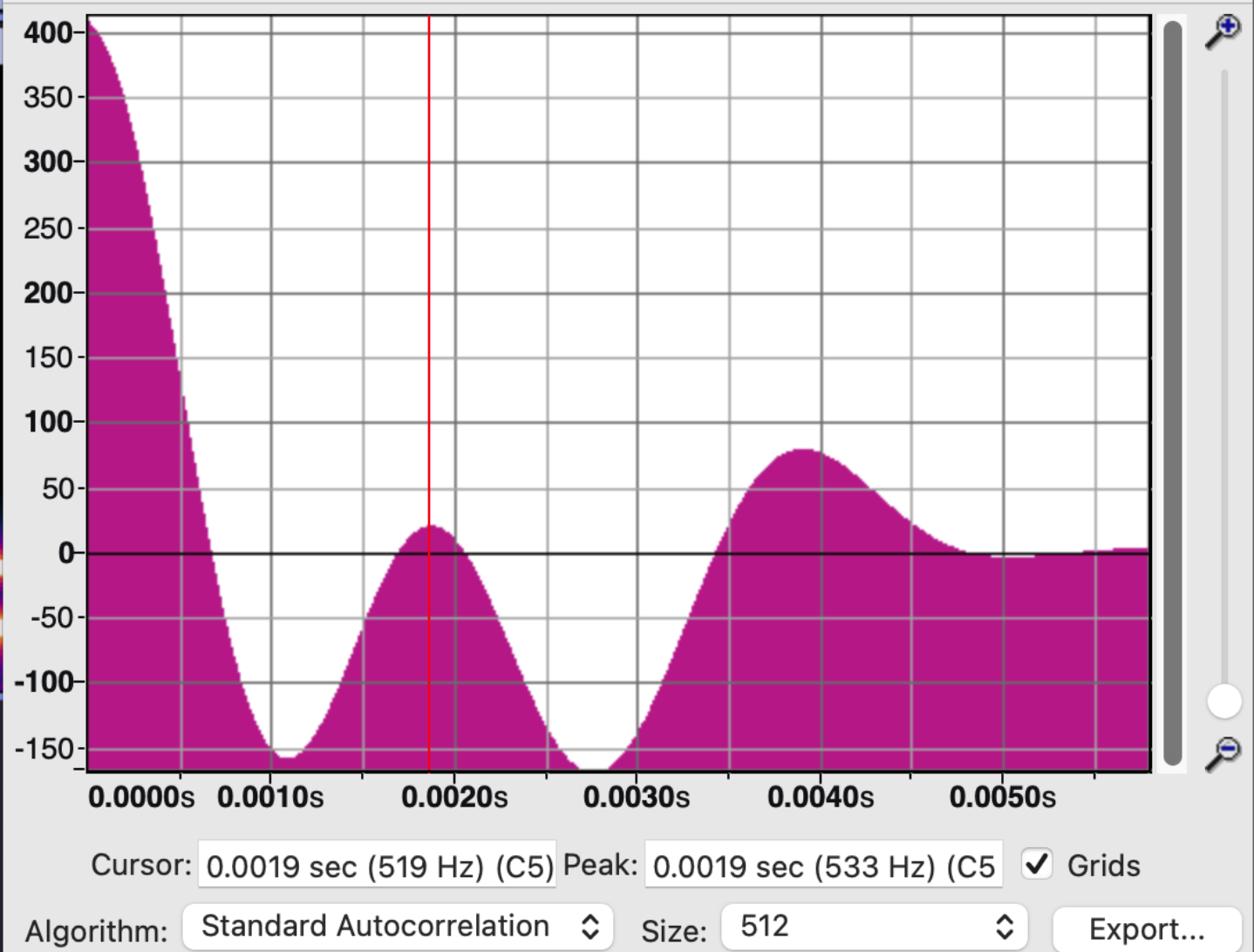}
  \end{minipage}
  \hfill
  \begin{minipage}[b]{0.48\textwidth}
    \centering
    \includegraphics[width=\textwidth]{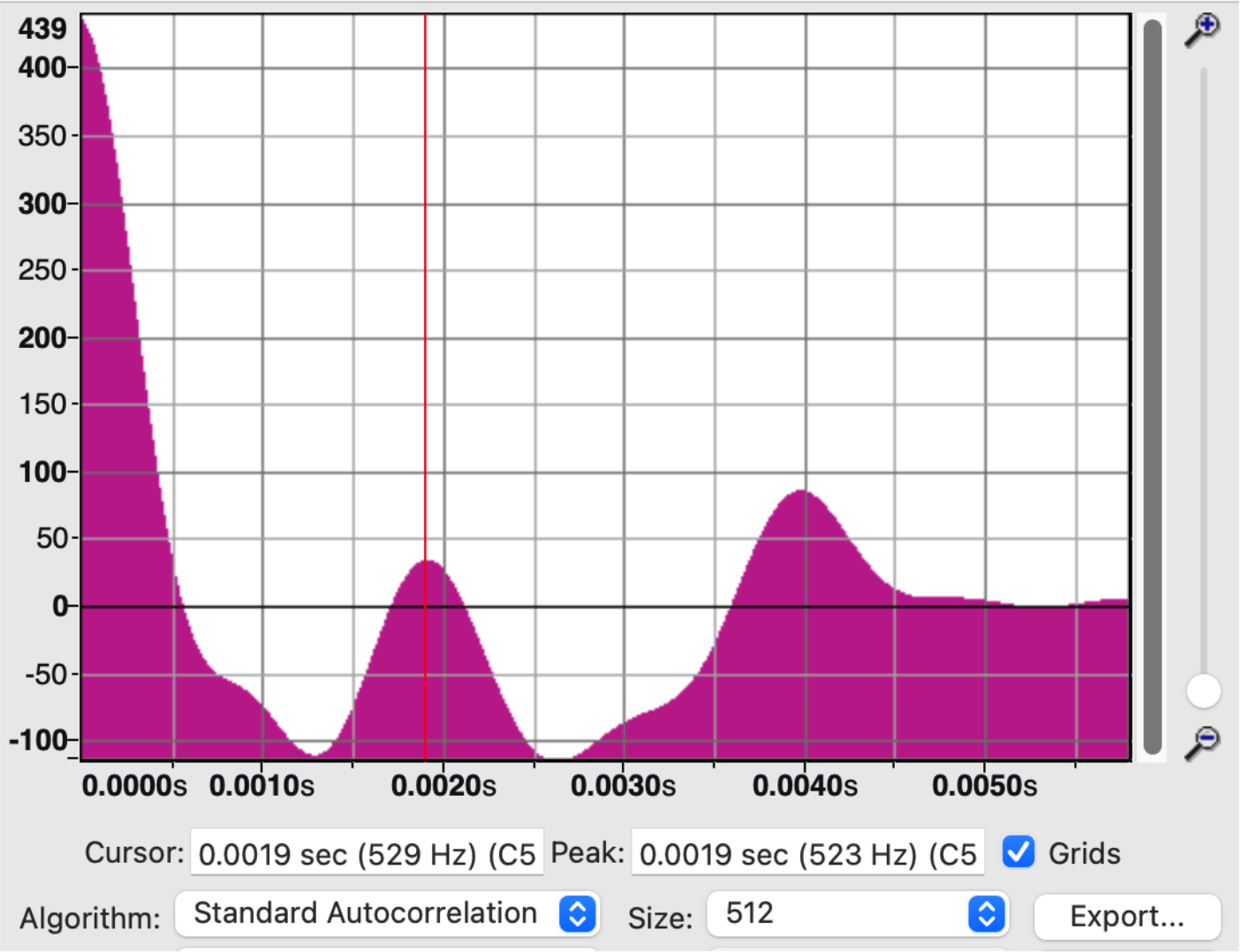}
  \end{minipage}
  \caption{Harmonic (left) and anharmonic (right) autocorrelations of water, scaled from a fundamental sine frequency of 220 Hz. Unlike HCl, there are two prominent peaks present, where the one that arrives later has slightly greater height, corresponding to a famous auditory ambiguity in autocorrelation functions.}
  \label{ch02:fig06}
\end{figure}

The autocorrelations (Fig.~\ref{ch02:fig06}) were of particular interest because they actually hint towards a perceptual ambiguity in psychoacoustics. There are two peaks, one at 533 Hz and 256 Hz for harmonic water and 523 Hz and 251 Hz for anharmonic water. One can actually listen for the prevalence of one or the other in the sound file.

While the inverse time of the earliest, most prominent peak of the autocorrelation function is related to the perceived pitch, there are cases when the peaks are of the same height, or one peak is larger than the other but arrives later. This is the idea behind Diana Deutsch's famous tritone paradox \citep{deutsch_tritone_1991}. Despite this intriguing "landscape" of the autocorrelation function, we do find that the anharmonic molecule is consistently "flatter" in pitch than the harmonic, which correlates with our predictions based on \eqref{eq:anharmonic}.

\subsection{Polyatomics}
While we also analyzed methanol and acetone spectra, for brevity and length we will simply talk about ammonia, but will leave the sound files and figures for viewing on the supplemental website. While water was computationally manageable due to only having three vibrational modes, the polyatomic molecules generally required heuristic approximations and decisions as to which combinations of energy modes are less likely to be audible or prominent in the spectrum, as briefly outlined in Methods. 

\begin{figure}[!h]\centering
  \begin{minipage}[b]{0.48\textwidth}
    \centering
    \includegraphics[width=\textwidth]{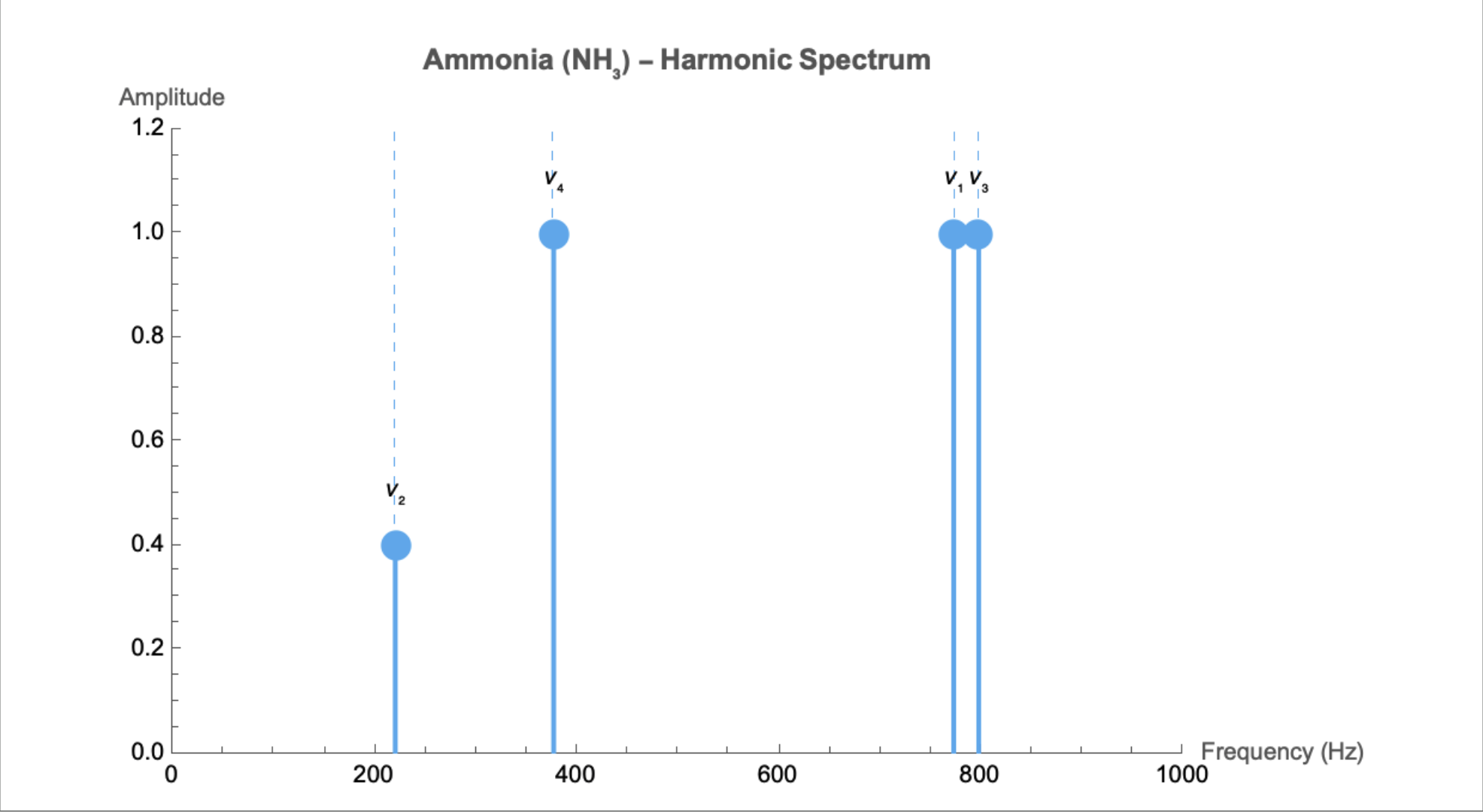}
  \end{minipage}
  \hfill
  \begin{minipage}[b]{0.48\textwidth}
    \centering
    \includegraphics[width=\textwidth]{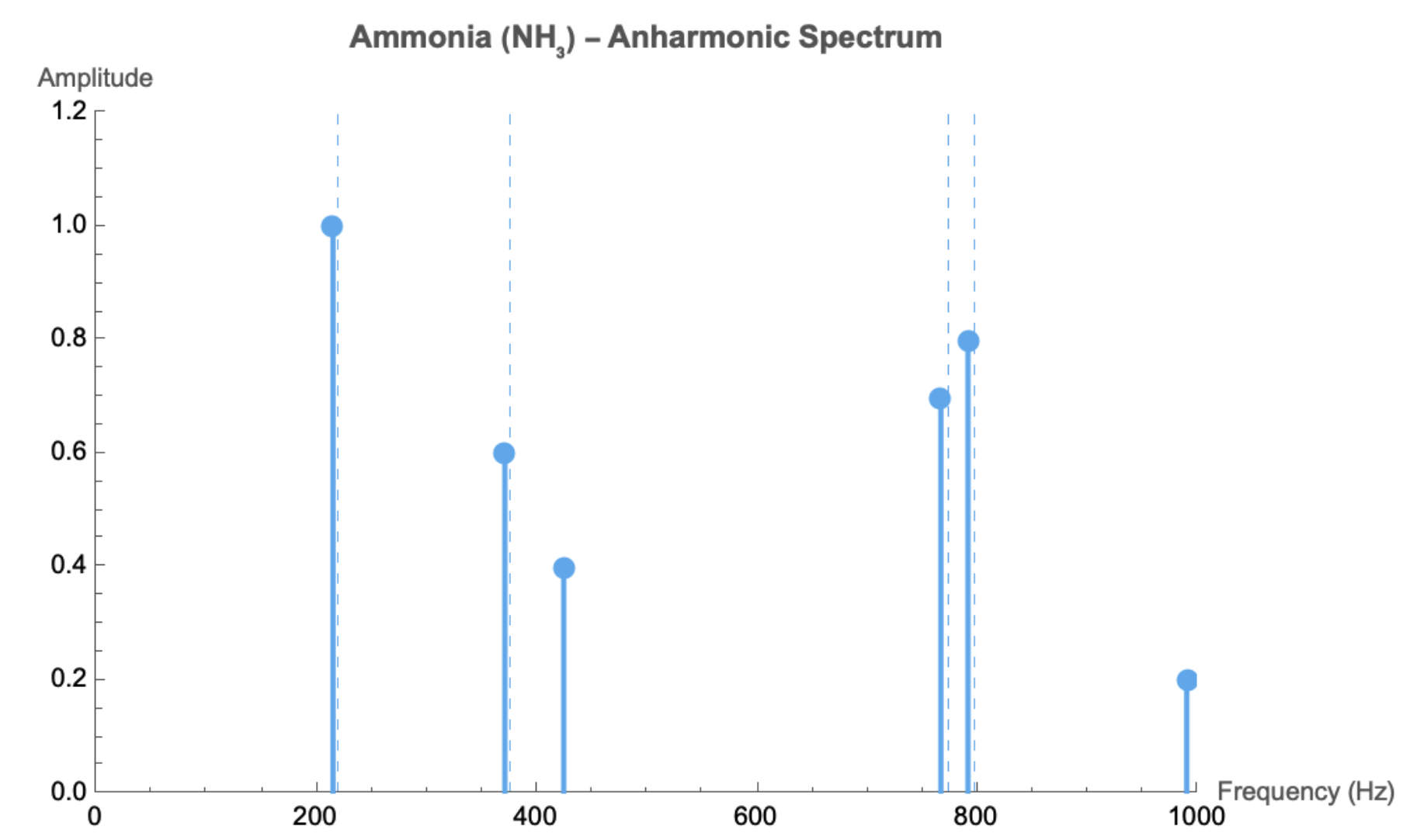}
  \end{minipage}
  \caption{Harmonic (left) and anharmonic (right) spectra of ammonia, scaled from a fundamental sine frequency of 220 Hz. Less prominent combinations of energy modes have been omitted for computational efficiency.}
  \label{ch02:fig07}
\end{figure}

\begin{figure}[!h]\centering
  \begin{minipage}[b]{1\textwidth}
    \centering
    \includegraphics[width=\textwidth]{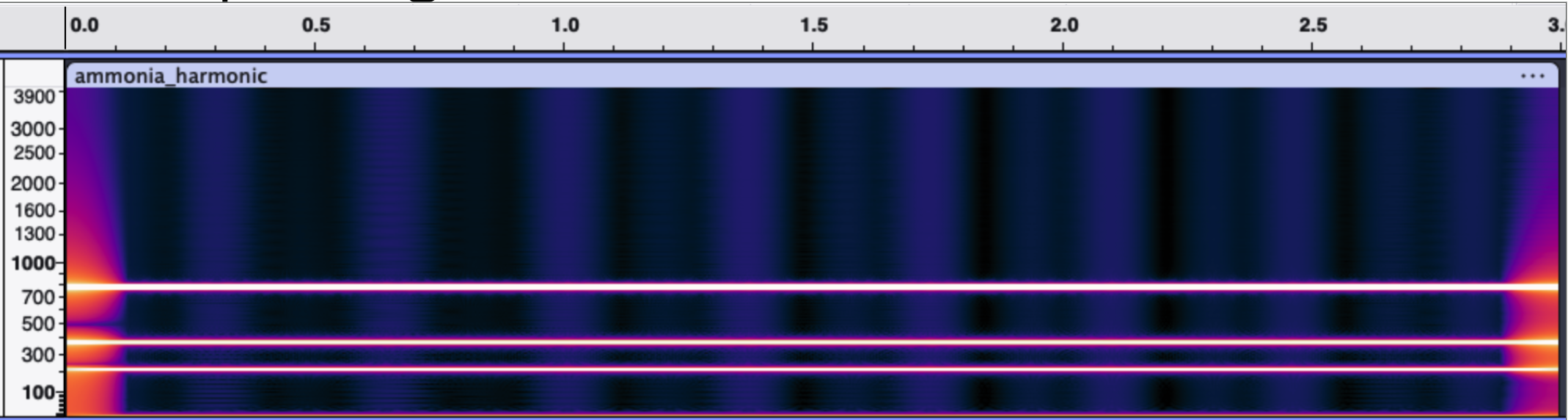}
  \end{minipage}
  \hfill
  \begin{minipage}[b]{1\textwidth}
    \centering
    \includegraphics[width=\textwidth]{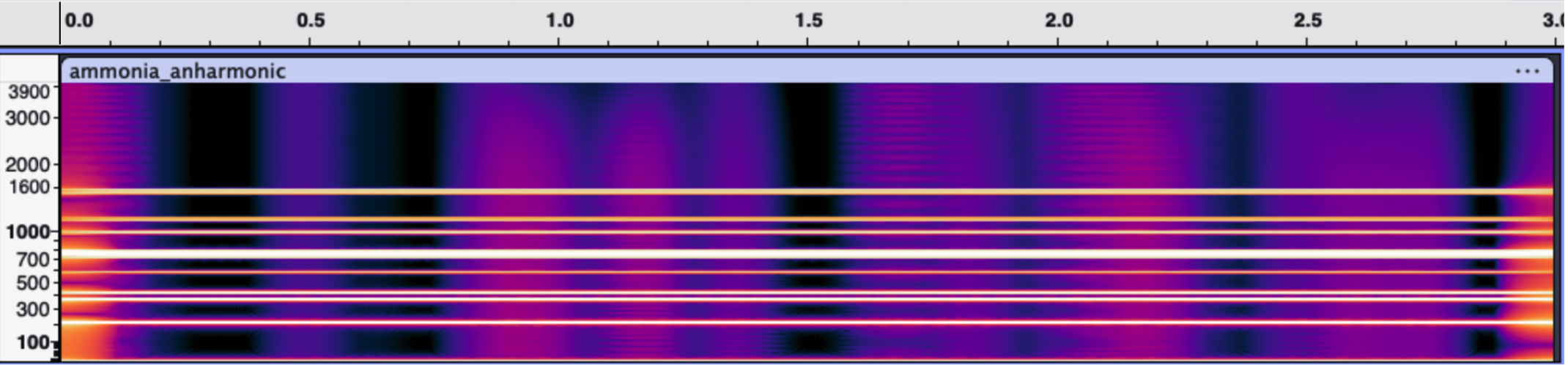}
  \end{minipage}
  \caption{Harmonic (top) and anharmonic (bottom) spectrograms of ammonia, scaled from a fundamental sine frequency of 220 Hz. The anharmonic spectrogram shows more complexity and also brighter vertical bands than in the harmonic.}
  \label{ch02:fig08}
\end{figure}

Similarly to water, the base frequency for ammonia was 220 Hz. There are four vibrational modes taken and scaled from NIST data \citep{nist_webbook}, and the spectra are shown in Fig.~\ref{ch02:fig07}. As before, each vibrational mode is slightly left-shifted and there are some combination of energy modes present, with weaker or less prominent ones omitted for computational efficiency. The spectrograms were also analyzed of these sound files (Fig.~\ref{ch02:fig08}). 

\begin{figure}[!h]\centering
  \begin{minipage}[b]{0.48\textwidth}
    \centering
    \includegraphics[width=\textwidth]{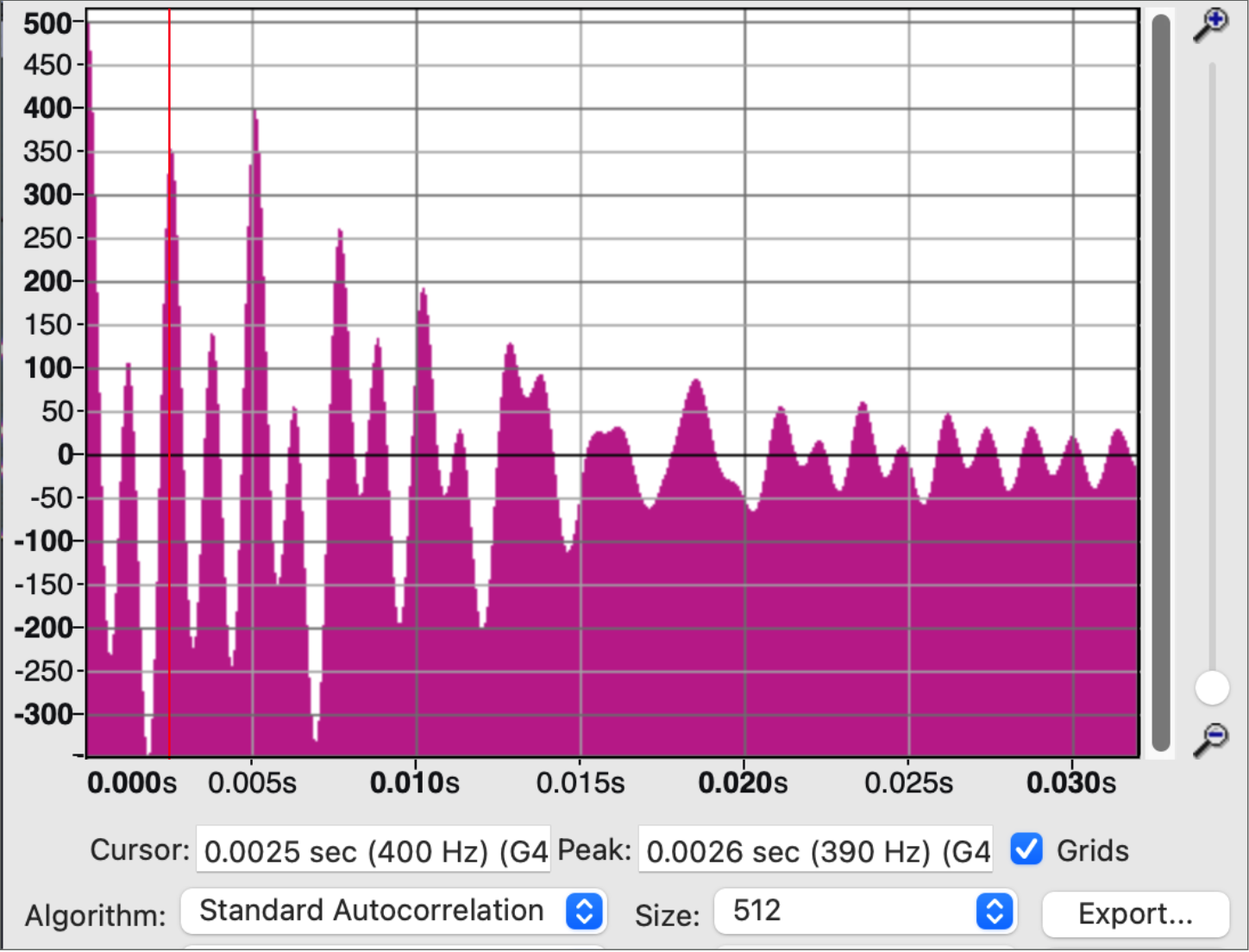}
  \end{minipage}
  \hfill
  \begin{minipage}[b]{0.48\textwidth}
    \centering
    \includegraphics[width=\textwidth]{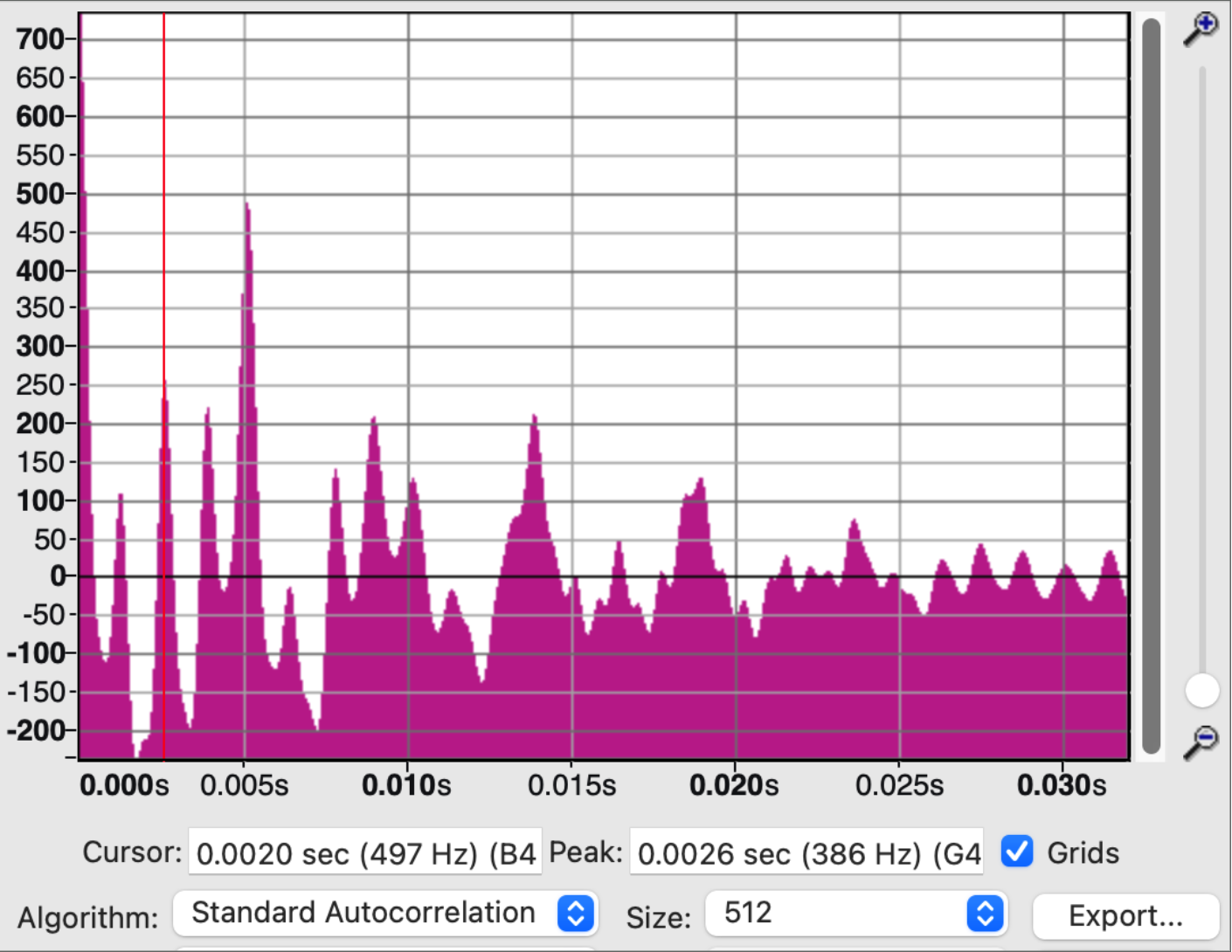}
  \end{minipage}
  \caption{Harmonic (left) and anharmonic (right) autocorrelations of ammonia, scaled from a fundamental sine frequency of 220 Hz. There are many jagged peaks, adding to more ambiguity in perceived pitch.}
  \label{ch02:fig09}
\end{figure}

The spectrograms (Fig.~\ref{ch02:fig08}) show more complexity than the simpler water and diatomic molecules. This also entails increasing complexity along the vertical bands that appear in the spectrogram. They are fairly periodic in the harmonic case but lose their periodicity in the anharmonic case. The autocorrelation functions shown in Fig.~\ref{ch02:fig09} show perhaps a more extreme case of pitch perception ambiguity, with a jagged landscape of many competing peaks. 

\subsection{Intramolecular Vibrational Energy Redistribution (IVR)}

We modeled IVR with acetone, which is not too complicated but still an interesting polyatomic molecule of interest. We are specifically plucking the C=O double bond. The Mathematica code to simulate IVR follows the derivation mentioned in Section 2.4. 

\begin{figure}
    \centering
    \includegraphics[width=0.8\linewidth]{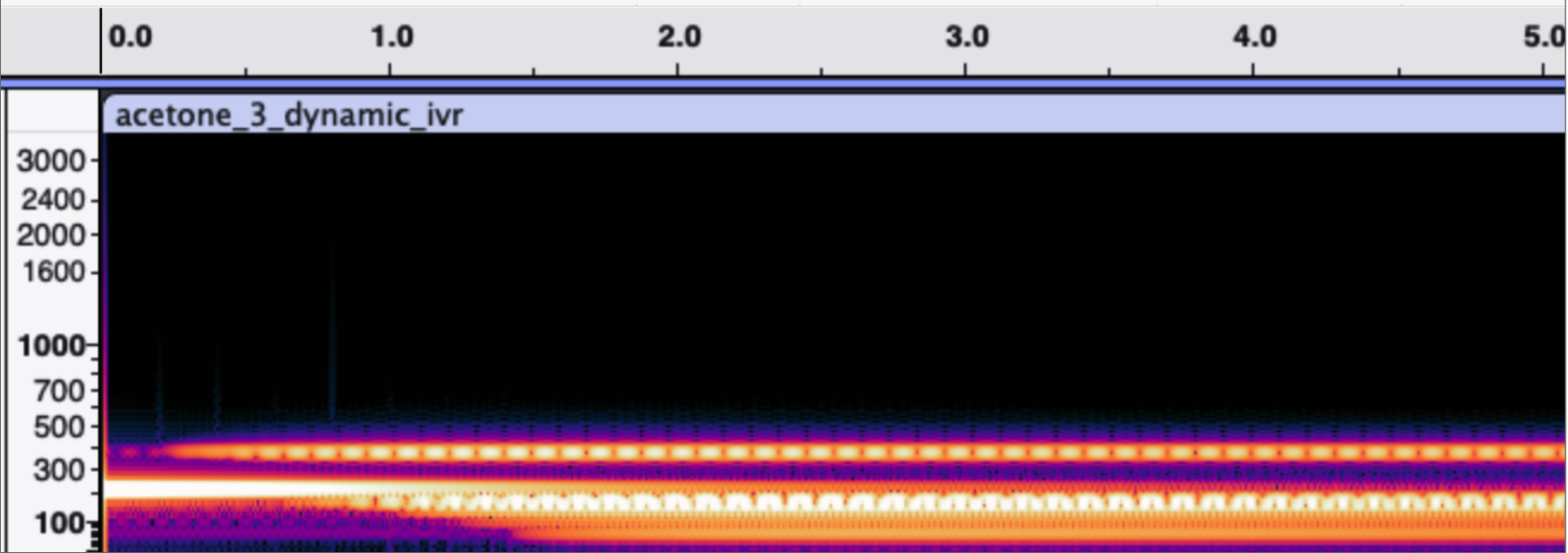}
    \caption{Spectrogram of IVR simulation of acetone. The first excited mode is most intense before decaying and exciting other neighboring vibrational modes on the molecule.}
    \label{ch02:fig10}
\end{figure}

The spectrogram below (Fig.~\ref{ch02:fig10}) shows a bright mode (C=O) that is initially excited before decaying. As it decays it transfers energies to the other modes that grow and wane in intensity. 

\begin{figure}[!h]\centering
  \begin{minipage}[b]{0.48\textwidth}
    \centering
    \includegraphics[width=\textwidth]{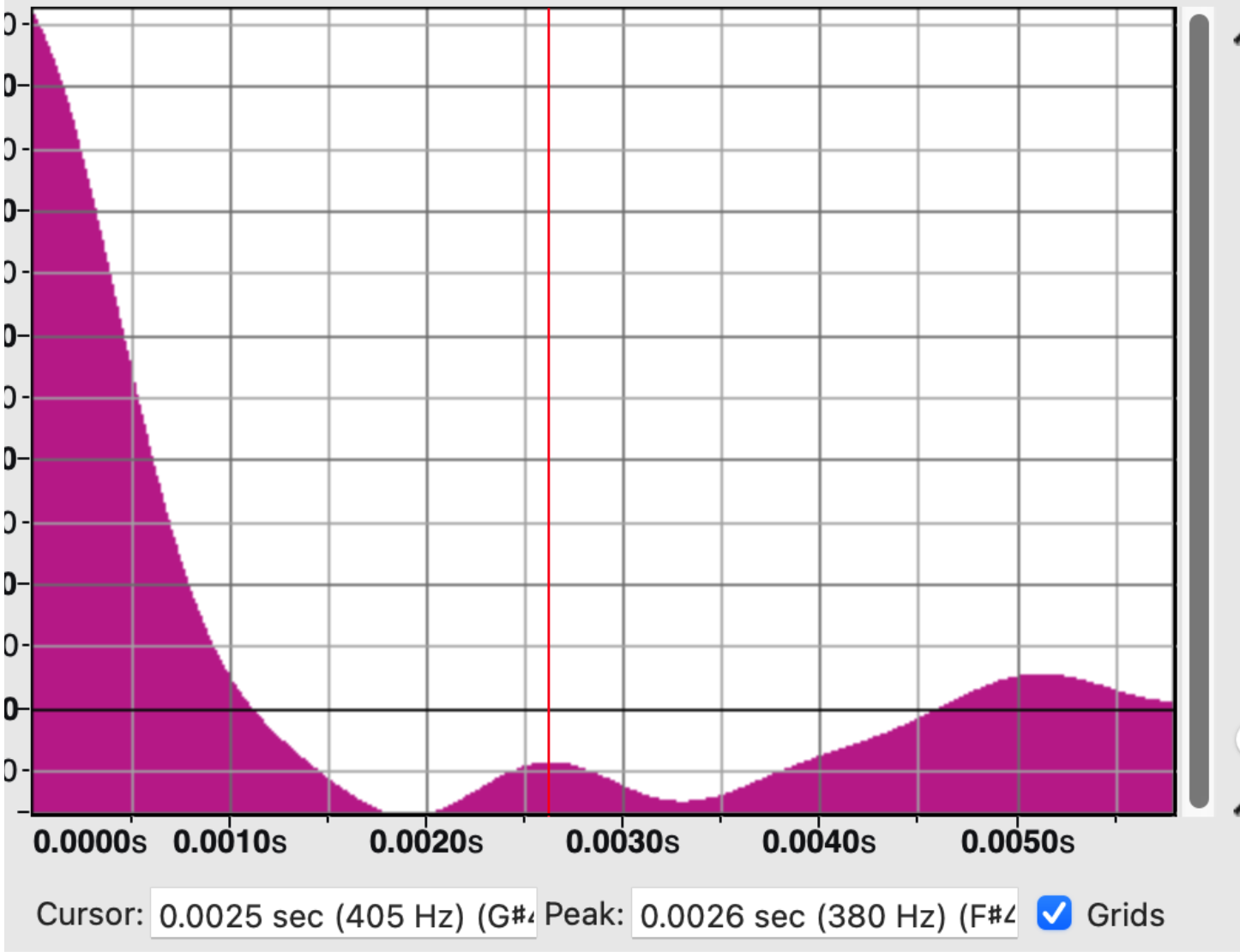}
  \end{minipage}
  \hfill
  \begin{minipage}[b]{0.48\textwidth}
    \centering
    \includegraphics[width=\textwidth]{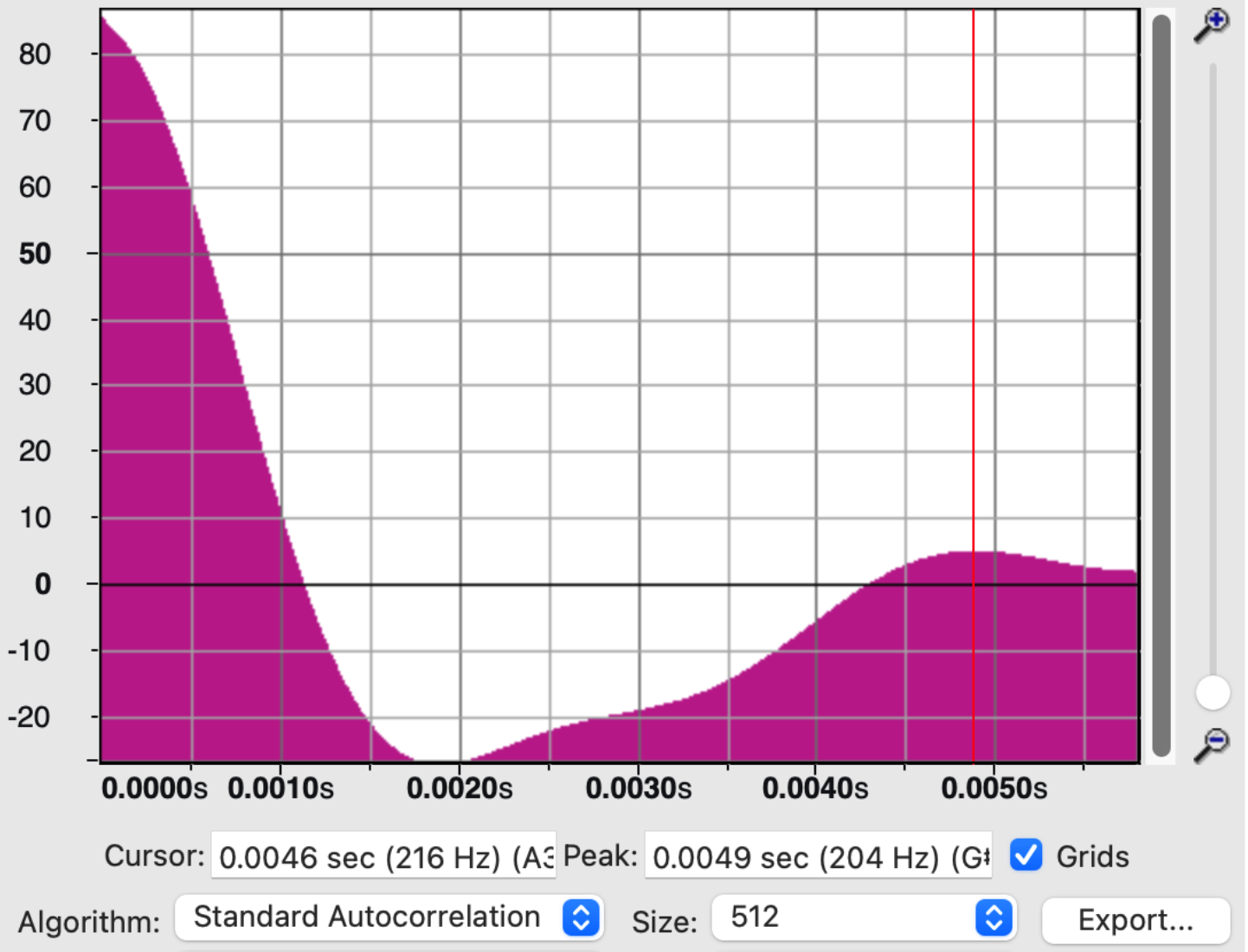}
  \end{minipage}
  \caption{Static (left) and IVR (right) autocorrelations of acetone. Most notably, what appears to have been the earlier of two peaks in the static plot on the left seems to have "merged" into the larger later peak in the right plot.}
  \label{ch02:fig11}
\end{figure}

The autocorrelation plots (Fig.~\ref{ch02:fig11}) are quite compelling as the "static" case (the anharmonic autocorrelation of acetone) has two competing peaks, similarly to water. However, the shorter earlier peak appears to "blend" into the later larger peak in the "dynamic" or IVR autocorrelation plot. The IVR simulation is the first instance of a time-dependent sound signal with timed excitations and decays of energy modes within the molecule, unlike a static combination of sine tones from previous simulations. 

\begin{figure}[!h]\centering
  \begin{minipage}[b]{0.75\textwidth}
    \centering
    \includegraphics[width=\textwidth]{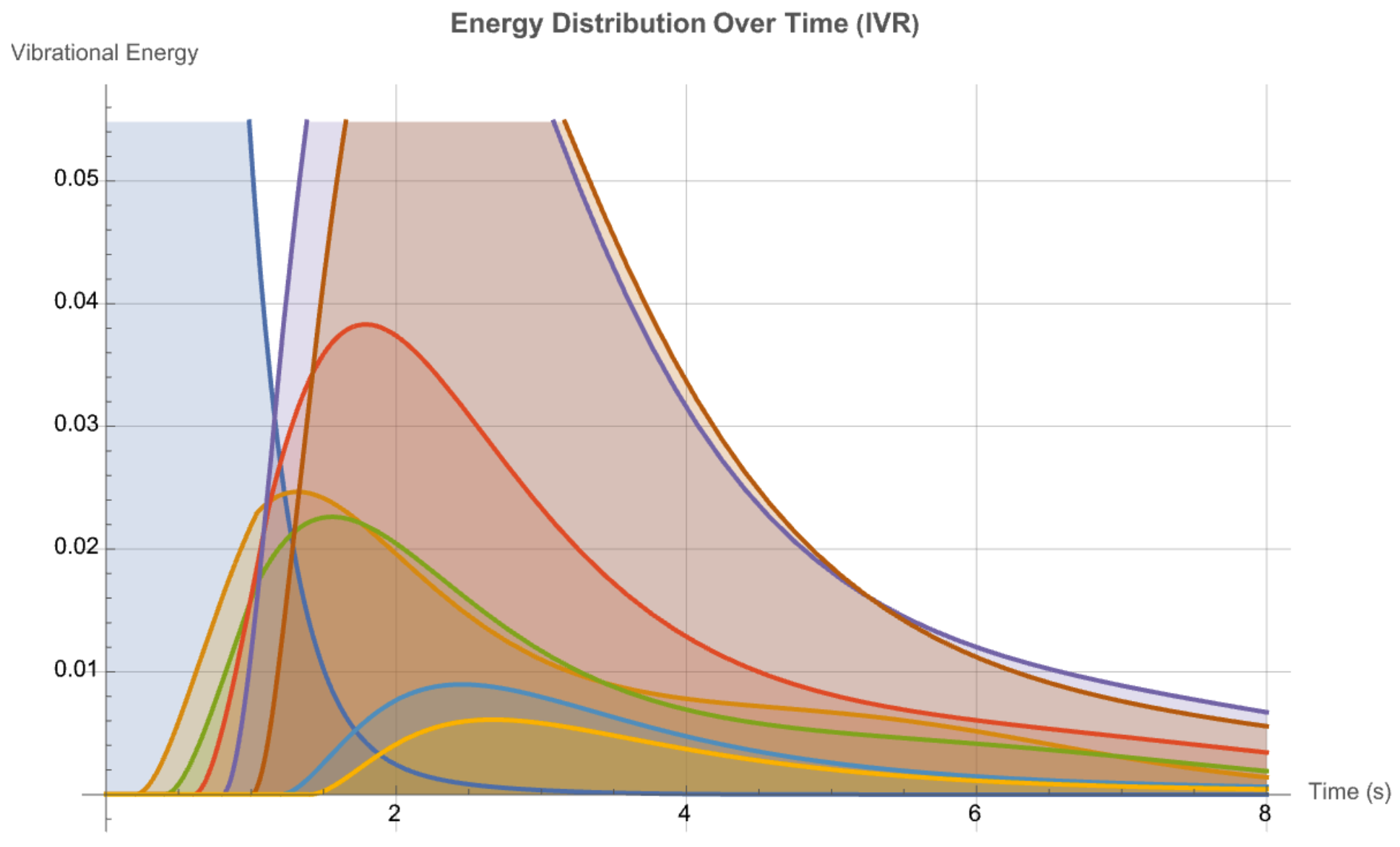}
  \end{minipage}
  \hfill
  \begin{minipage}[b]{0.20\textwidth}
    \centering
    \includegraphics[width=\textwidth]{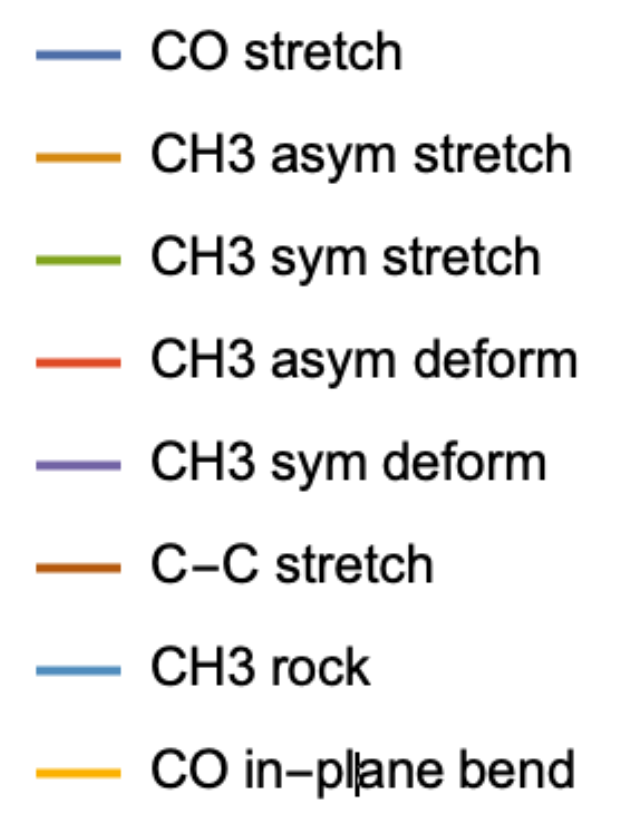}
  \end{minipage}
  \caption{Energy distribution over time of various energy modes, color-coded as depicted by legend on the right. The first excited mode, CO (blue) decays, immediately exciting CH3 asymmetric stretch and the other neighboring CH3 modes.}
  \label{ch02:fig12}
\end{figure}

To provide a better qualitative understanding of how the individual energy modes evolve over time, an energy distribution of time plot has also been plotted in Mathematica with a corresponding color scheme (Fig.~\ref{ch02:fig12}). The plots display how the first mode that has already been excited (C-O, in dark blue) decays quickly, subsequently exciting the neighboring $CH_3$ modes. As previously mentioned in Methods, the energy modes are modeled to exponentially decay. 

\section{Conclusion}

This paper provides an exploratory investigation into sonifying the IR spectra of small molecules, culminating in a time-dependent IVR simulation of acetone. Across all studied molecules, the leftward shifting of spectral peaks due to anharmonicity consistently causes the molecular "pitch" to flatten, as confirmed by autocorrelation analysis of the sound files. The inclusion of combination bands and overtones enriches the molecular "timbre," adding complexity beyond the fundamental modes. Not only do additional partials emerge in the anharmonic spectra, but distinctive beating patterns also arise from mode interference—patterns that are both audible and visible in spectrograms.

This work has several important limitations. The coupling matrices used for polyatomic molecules and IVR simulations were constructed heuristically based on general spectroscopic principles rather than calculated from first principles. Additionally, environmental effects such as solvent interactions and rotational-vibrational coupling were not considered. While the equations modeling IVR capture qualitative behavior, they may not compete with the rigor of developed IVR models in the literature \citep{Karmakar, semparithi_ivr_2006}.

Nevertheless, the paper achieves its central aim: providing an intuitive auditory experience of molecular vibrations. This exploration reveals unexpected connections to psychoacoustics, particularly the perceptual ambiguity that arises when competing peaks appear in an autocorrelation function. Listening to the supplementary sound files demonstrates how the ear is "forced" to resolve ambiguity by selecting one perceived pitch over another—a phenomenon familiar from musical perception but now observable in molecular spectra.

The IVR simulation of acetone illustrates how the autocorrelation "landscape" evolves dynamically. In the static anharmonic case, two competing autocorrelation peaks reflect the complex overtone structure. However, when time-dependent energy transfer is introduced through impulsive excitation of the C=O mode, one peak diminishes and merges into the other as energy redistributes throughout the molecule. This transformation suggests intriguing possibilities for applying psychoacoustical concepts to quantum chemistry: If we can pump molecules by exciting specific normal modes with narrowband laser light, could we instead pump at an autocorrelation-derived frequency? Would different outcomes result from targeting one autocorrelation peak versus another when multiple peaks compete? These questions remain open.

More broadly, there is exciting potential in applying psychoacoustical perspectives to the analysis of molecular dynamics. While molecular vibrations and musical sounds are not directly analogous systems, the parallels are sufficiently rich to warrant inspired investigation. Even if no deep theoretical connections emerge, sonification offers clear value for chemical education and outreach. Quantum chemistry is fundamentally mathematical and abstract, but being able to \textit{hear} vibrational dynamics may provide deeper conceptual intuition for students and researchers. At the very least, it can offer an aesthetically compelling way to engage with molecular structure. After all, the ability to "pluck" a molecule like a violin string surely adds a pleasant sensory experience to spectroscopic analysis.  

\bibliographystyle{unsrtnat}
\bibliography{references}  

@book{heller_semiclassical,
  author    = {Heller, Eric J.},
  title     = {The Semiclassical Way to Dynamics and Spectroscopy},
  publisher = {Princeton University Press},
  location  = {Princeton, NJ},
  year      = {2018},
}

@article{semparithi_ivr_2006,
  author  = {Semparithi, A. and Keshavamurthy, S.},
  title   = {Intramolecular Vibrational Energy Redistribution as State Space Diffusion: Classical--Quantum Correspondence},
  journal = {Journal of Chemical Physics},
  year    = {2006},
  volume  = {125},
  number  = {14},
  pages   = {141101},
  doi     = {10.1063/1.2358138},
}

@misc{audacity,
  author       = {{Audacity Team}},
  title        = {Audacity: Free, Open Source, Cross-Platform Audio Software},
  howpublished = {\url{https://www.audacityteam.org/}},
  note         = {Version 3.5.1, accessed 19~Dec.~2025},
  year         = {2025},
}

@Article{Karmakar,
author ="Karmakar, Sourav and Keshavamurthy, Srihari",
title  ="Intramolecular vibrational energy redistribution and the quantum ergodicity transition: a phase space perspective",
journal  ="Phys. Chem. Chem. Phys.",
year  ="2020",
volume  ="22",
issue  ="20",
pages  ="11139-11173",
publisher  ="The Royal Society of Chemistry",
doi  ="10.1039/D0CP01413C",
url  ="http://dx.doi.org/10.1039/D0CP01413C"}

@article{deutsch_tritone_1991,
  author  = {Deutsch, Diana},
  title   = {The Tritone Paradox: An Influence of Language on Music Perception},
  journal = {Music Perception},
  year    = {1991},
  volume  = {8},
  number  = {4},
  pages   = {335--347},
}

@Article{Arasaki,
author ="Arasaki, Yasuki and Takatsuka, Kazuo",
title  ="Sonification of molecular electronic energy density and its dynamics",
journal  ="RSC Adv.",
year  ="2024",
volume  ="14",
issue  ="13",
pages  ="9099-9108",
publisher  ="The Royal Society of Chemistry",
doi  ="10.1039/D4RA00999A",
url  ="http://dx.doi.org/10.1039/D4RA00999A"}

@article{mahjour_molecular_2023,
  author  = {Mahjour, Babak and Bench, Jordan and Zhang, Rui and Frazier, Jared and Cernak, Tim},
  title   = {Molecular Sonification for Molecule to Music Information Transfer},
  journal = {Digital Discovery},
  year    = {2023},
  volume  = {2},
  pages   = {520--530},
  doi     = {10.1039/D3DD00008G},
}

@misc{nist_webbook,
  author       = {Linstrom, Peter J. and Mallard, William G.},
  title        = {NIST Chemistry WebBook, NIST Standard Reference Database Number 69},
  howpublished = {\url{https://webbook.nist.gov/chemistry/}},
  note         = {National Institute of Standards and Technology, Gaithersburg, MD},
  year         = {1998},
}

@misc{anharmonic_oscillator_notes,
  title        = {Notes on Quantum Field Theory: 7 {Anharmonic} Oscillator},
  howpublished = {\url{https://noteimages.s3.amazonaws.com/anharmonic_oscillator.pdf}},
  note         = {Lecture notes, accessed 19~Dec.~2025},
}

@article{Pereira,
author = {Pereira, Florbela and Ponte-e-Sousa, João C. and Fartaria, Rui P. S. and Bonifácio, Vasco D. B. and Mata, Paulina and Aires-de-Sousa, Joao and Lobo, Ana M.},
title = {Sonified Infrared Spectra and Their Interpretation by Blind and Visually Impaired Students},
journal = {Journal of Chemical Education},
volume = {90},
number = {8},
pages = {1028-1031},
year = {2013},
doi = {10.1021/ed4000124},

URL = { 
    
        https://doi.org/10.1021/ed4000124
    
    

},
eprint = { 
    
        https://doi.org/10.1021/ed4000124
    
    

}

}

@article{Garrido,
author = {Garrido, Neil and Pitto-Barry, Anaïs and Soldevila-Barreda, Joan J. and Lupan, Alexandru and Boyes, Louise Comerford and Martin, William H. C. and Barry, Nicolas P. E.},
title = {The Sound of Chemistry: Translating Infrared Wavenumbers into Musical Notes},
journal = {Journal of Chemical Education},
volume = {97},
number = {3},
pages = {703-709},
year = {2020},
doi = {10.1021/acs.jchemed.9b00775},

URL = { 
    
        https://doi.org/10.1021/acs.jchemed.9b00775
    
    

},
eprint = { 
    
        https://doi.org/10.1021/acs.jchemed.9b00775
    
    

}

}






\end{document}